\documentclass[a4paper]{jpconf}
\usepackage{graphicx}
\usepackage{amsmath,amssymb}
\usepackage{aas_macros}
\begin{document}
\title{Radiative Models of Sagittarius A* and M87 from
  Relativistic MHD Simulations}

\author{J. Dexter$^{1}$, E. Agol$^{2}$, P. C. Fragile$^{3}$ and J. C. McKinney$^{4}$}

\address{$^{1}$Theoretical Astrophysics Center and Department of Astronomy,
  University of California, Berkeley, CA 94720-3411, USA}
\address{$^{2}$Department of Astronomy, University of Washington, Box 351580, Seattle, WA 98195, USA}
\address{$^{3}$Department of Physics \& Astronomy, College of Charleston, Charleston, SC 29424, USA}
\address{$^{4}$Kavli Institute for Particle Astrophysics and Cosmology, Stanford University, Stanford, CA 94305-4060, USA}

\ead{jdexter@berkeley.edu}

\begin{abstract}
Ongoing millimeter VLBI observations with the Event Horizon Telescope
allow unprecedented study of the innermost portion of black hole
accretion flows. Interpreting the observations requires relativistic,
time-dependent physical modeling. We discuss the comparison of radiative
transfer calculations from general 
relativistic MHD simulations of Sagittarius A* and M87 with current
and future mm-VLBI observations. This
comparison allows estimates of the viewing geometry and physical
conditions of the Sgr A* accretion flow. The viewing geometry for M87
is already constrained from observations of its large-scale jet, but,
unlike Sgr A*, there is no consensus for its millimeter emission
geometry or electron population. Despite this uncertainty, as long as
the emission region is compact, robust predictions for the size of its
jet launching region can be made. For both sources, the black hole shadow may be detected with 
future observations including ALMA and/or the LMT, which would
constitute the first direct evidence for a black hole event horizon.
\end{abstract}


\section{Introduction}

The Galactic center massive black hole candidate, Sagittarius A* (Sgr
A*) \cite{balick1974}, and the supermassive black hole candidate in
the center of M87 are both bright radio sources and are the two
largest black holes in angular size ($4-5$ microarcseconds or
$\mu$as). For both these reasons, they are the two major targets of
interest for imaging with ongoing very long
baseline interferometry observations at millimeter wavelengths
(mm-VLBI), which have already detected time-variable event horizon
scale structure in Sgr A* \cite{doeleman2008,fishetal2011}. Data have
also been taken for M87, but are as yet unpublished. The Sgr A* data have
previously been fit both with simple geometric models (symmetric
Gaussian and annulus of constant brightness) and using a radiatively
inefficient accretion flow model (RIAF,
\cite{yuanquataert2003,broderick2009,brodericketal2011}. While physically motivated,
these models are non-relativistic and do not include the magnetic
fields that drive turbulence in the accretion flow via the
magnetorotational instability (MRI, \cite{mri}), causing accretion and
leading to the observed synchrotron radiation. 

This essential physics can now be captured in general relativistic MHD
(GRMHD) simulations of black hole accretion. Simulations have 
been used to model the millimeter emission in Sgr A* either
in the absence of general relativity \cite{goldston2005} or in
axisymmetry \cite{noble2007,moscibrodzka2009,shcherbakovetal2011}. The
millimeter emission 
in the models arises in the innermost portions of the accretion flow
($r \simeq 5$M, where we use $G=c=1$ units except where otherwise
noted), so that relativistic effects are dominant. In simulations run
in axisymmetry, the MRI decays on the local orbital timescale. This
leads to unrealistic variability, prevents the disk from reaching
quasi-steady state, and can cause the appearance of unphysically hot
and bright regions \cite{moscibrodzka2009}. One group
\cite{shcherbakovetal2011} has used 3D GRMHD simulations, but then
time-averaged the resulting data. Two groups have compared spectra
from 2D GRMHD simulations with the spectrum of M87
\cite{moscibrodzkaetal2011,hilburnliang2011}, but neither considered
jet emission or made predictions for mm-VLBI.

Here we discuss radiative models of the mm emission from Sgr A* and
M87 for comparison with mm-VLBI constructed by performing relativistic
radiative transfer calculations on 3D GRMHD simulation data. These are the first
models to self-consistently include variability, and the M87 models
are the first predictions for images of a jet launching region from a
simulation. The numerical models of Sgr A* are the only ones to have
been fit directly to mm-VLBI data to test GRMHD accretion theory and
estimate parameters of the black hole and its accretion flow.

\section{Radiative Modeling}

A radiative model of Sagittarius A* or M87 consists of three
components \cite{dexter2011}: a dynamical solution for the fluid variables
in the accretion flow, a model for the electron distribution function
based on the dynamical solution, and an emission model given the
electron distribution function. These components give fluid frame
emission/absorption coefficients everywhere in spacetime. Observables
are calculated from these coefficients in full general relativity via
ray tracing \cite{luminet1979}.

\subsection{GRMHD Simulations}

The dynamical solution consists of the output of general relativistic MHD
simulations of black hole accretion flows. The simulations start from a gas torus
in hydrostatic equilibrium in a Kerr spacetime threaded with a weak
magnetic field. This configuration quickly develops turbulence driven
by the MRI, and the resulting stresses cause angular momentum
transport outwards allowing accretion onto the black hole. These
simulations are currently limited in several 
important respects: they only develop
a quasi-steady accretion flow over a limited radial domain ($r 
\lesssim 25$M); and they neglect both the radiation that astronomers
observe and the electrons that produce it.

In highly underluminous systems like Sgr A* and M87, ignoring the
effects of radiation on the dynamical solution is reasonable (but see
\cite{drappeauetal2011} and 
Dibi et al., in prep.). The limited radial domain is also acceptable 
given the extremely compact size measured by mm-VLBI, and found
theoretically for the millimeter emission in Sgr A* and M87. 

However, there are also issues with modeling these sources using GRMHD
simulations. Their low densities means that the ions and electrons may
have significantly different temperatures, with the hot ions providing the
fluid pressure captured by the simulation and the cool electrons
producing the observed emission. We must then decide how to model the
electrons before calculating observables from simulation data (see
\S~\ref{sec:electrons}). 

In this work, we consider results from four GRMHD simulations
(summarized in Table 1 of \cite{dexteretal2010}). All are
three dimensional: in axisymmetry, the MRI decays on the local orbital
timescale, preventing the accretion flow from reaching quasi-steady
state and leading to artificial variability, an inaccurate
time-averaged structure, and causing contamination of the spectra by
unrealistically hot zones in the ``funnel wall''
\cite{moscibrodzka2009}. The simulations are from two codes: 
\texttt{Cosmos++} \cite{anninos2005,fragile2007,fragileetal2009} and a
3D extension of HARM
\cite{gammie2003,noble2006,mckinney2006ff,tchekhovskoyetal2007,mckinneyblandford2009}. 
The set of simulations is limited by the number of existing and available
3D GRMHD simulations of relatively thick ($H/R \simeq 0.2$) black hole
accretion flows. The scarcity of simulations and their systematic
differences prevent an estimate of the black hole spin. Only the MBD 
simulation forms a relativistic, magnetically dominated-jet and so
this is the only simulation used in the disk/jet modeling of M87. We use the last
$2000-4000$M of simulation data, once transients from the initial
conditions have decayed,  where $1$M is 
$20$s ($9$hr), $6\times10^{11}$cm ($10^{15}$cm), and $5\mu$as
($4\mu$as) for Sgr A* (M87). 

\subsection{Electrons}\label{sec:electrons}

The millimeter emission in Sgr A* is well described by synchrotron
emission from hot, thermal electrons
in the innermost portions of a radiatively inefficient accretion flow
\cite{narayanyi1994,yuanquataert2003}. A non-thermal tail extending to
higher energies may be responsible for the radio and IR emission
\cite{yuanquataert2003}, or it may arise in a
jet \cite{falckemarkoff2000}. 

Since we focus on the millimeter emission, we model accretion disk
components as having 
entirely thermal electrons. The electron temperature is specified
using a constant electron-ion temperature ratio, $T_i/T_e$
\cite{goldston2005,moscibrodzka2009}, which 
allows the electron temperature to be calculated from the simulation 
pressure and density. This is the simplest possible prescription for a
two-temperature accretion flow from a single fluid simulation.


In M87, it is unclear whether disk or jet emission is dominant at millimeter
wavelengths. Extended jet structure
is observed at 7mm \cite{lyetal2004,walkeretal2008}, indicating that the jet emission is
at least comparable to that from the disk. Since in the simulations
the jet forms self-consistently from an accretion flow, we include
both disk and jet components with different assumed electron
populations \cite{dexteretal2012}.

The ultrarelativistic outflow in the simulation forms in
magnetically-dominated region, so to separate the regions for electron
modeling we define the jet by $b^2/\rho c^2 > 1$, where $b$ is the magnetic
field strength and $\rho$ is the fluid density. The jet electrons are
assumed to be entirely non-thermal with a power law distribution. The
simulation variables of density and internal energy (pressure) within
magnetically-dominated regions are unreliable, since they are set by
the artificial numerical floor values used for code stability. We therefore scale the
non-thermal particle density with the only reliable simulation
variable in the jet region, the magnetic energy density: $n_{\rm nth}
\propto b^2$, where $n_{\rm nth}$ is the non-thermal particle
density (cf. \cite{broderickmckinney2010}). Fitting the M87 spectrum requires $5-10\%$ of the magnetic
energy to be dissipated in non-thermal particles for the simulation
considered. 


In both cases, the emissivity is assumed to be entirely synchrotron
radiation. Inverse Compton scattering is ignored, although it is likely to be
very important for M87 at high energies
\cite{moscibrodzkaetal2011,dexteretal2012}. For the thermal disk
electrons, we use an approximate form of the synchrotron emissivity given by Leung et al. (2011)
\cite{leungetal2011}. For the jet power law electron distribution, we use a form
of the emissivity that takes into account the low-energy
cutoff to the distribution, which is important for
the millimeter emission in M87 \cite{dexter2011}.



\subsection{Ray Tracing}

Relativistic radiative transfer is performed on the simulation data
via ray tracing using the code \texttt{grtrans}
\cite{dexter2011}. Starting from an observer's camera, rays are
traced backwards in time toward the black hole assuming they are null
geodesics (geometric optics approximation), using the public code
\texttt{geokerr} \cite{dexteragol2009}. In the region where rays
intersect the accretion flow, the radiative transfer equation is solved along the
geodesic, which then represents a pixel of the image. This
procedure is repeated for many rays to produce an image, and at many
time steps of the simulation to produce time-dependent images
(movies). Light curves are computed by integrating over the individual
images. Repeating the procedure over observed wavelengths gives a
time-dependent spectrum.

To calculate fluid properties at each point on a ray, the spacetime
coordinates of the geodesic are transformed from Boyer-Lindquist to
the regular \cite{fragile2005} or modified \cite{mckinney2006}
Kerr-Schild coordinates used in the simulation. Since the accretion
flow is dynamic, light travel time delays along the geodesic are taken
into account. Data from the sixteen nearest zone centres (eight on the
simulation grid over two time steps) are interpolated to each point
on the geodesic.

Computing emission and absorption coefficients requires converting
simulation fluid variables (pressure/internal energy, mass density,
and magnetic field strength) into an electron distribution function in
physical units. The black hole mass sets the length and time scales,
while the mass of the initial torus provides an independent scale and
fixes the accretion rate. The scalings are such that $n$ and $b^2$ are
proportional to the accretion rate.

\section{Model Fitting}

For Sgr A*, we fit simulated images at $1.3$mm to mm-VLBI observations
\cite{doeleman2008,fishetal2011} 
and the total flux at $1.3$mm and $0.4$mm to observations of the
sub-mm spectral index \cite{marronephd}. Details on the statistical
techniques used to estimate parameters from the fits can be found in
\cite{broderick2009,dexteretal2010}. The free parameters for the Sgr
A* models are i) the time-averaged accretion rate, ii) the ratio
$T_i/T_e$, iii) the inclination and iv) sky orientation angles, and v)
the simulation.

Generically, the simulated images provide excellent fits to the
mm-wavelength observations of Sgr A* (reduced $\chi^2 \lesssim 1$ for
many models for each simulation
\cite{dexter2009,dexteretal2010}). Probability distributions for the
inclination ($i$) and sky orientation ($\xi$) 
angles, the median electron temperature in the region that produces
$99\%$ of the millimeter emission ($T_e$), and the time-averaged accretion
rate onto the black hole ($\dot{M}$) including all mm-VLBI data are estimated 
as follows (all to $90\%$ confidence): $i=60\pm15^\circ$,
$\xi={-70^\circ}^{+86^\circ}_{-15^\circ}$, $T_e=6\pm2 \times 10^{10}$K
and $\dot{M}=3^{+7}_{-1} \times 10^{-9} M_\odot \mathrm{yr}^{-1}$. Including all of the
data gives tighter constraints, but consistent 
estimates, for all parameters as found from the first epoch alone. The viewing angle estimates and
probability distributions are in excellent agreement with those from
using a RIAF model
\cite{broderick2009,brodericketal2011}. The mm-VLBI observations, even
with limited array coverage and sensitivity, are meaningfully
constraining the relevant parameters, but they cannot yet distinguish
between physical accretion flow models (they are, however, beginning to favor
physical models over simple geometric ones
\cite{brodericketal2011}). The constraints on the accretion rate and
electron temperature are consistent with, but independent of,
estimates from the observed linear polarization and Faraday rotation
measures \cite{agol2000,quataertgruzinov2000,marrone2007} and spectral
fitting \cite{yuanquataert2003}, respectively. 

The first mm-VLBI data for M87 have been obtained \cite{fishetal2011}
but have not yet been published. We fit models to multi-epoch
observational data in the optical \cite{sparksetal1996}, IR 
\cite{perlmanetal2001,perlmanetal2007}, and mm 
\cite{tanetal2008}, inflating the error bars to account
for variability between the non-simultaneous observations and treating
data points that may be dominated by dust in the galactic nucleus or
large-scale jet emission as upper limits. Since there are roughly the
same number of free parameters in the disk/jet models as spectral
constraints, we identify
fiducial models 
rather than attempt to quantiatively constrain the parameters. The
models can be either disk/jet (DJ1) or jet only (J2), and provide
satisfactory descriptions of the emission from the immediate vicinity
of the M87 black hole \cite{dexteretal2012}.

The viewing geometry of M87, unlike Sgr A*, may be constrained by
the Lorentz factor and sky orientation of the large-scale jet, which
can reasonably be assumed to correspond to the black hole spin
axis. All models assume this favored geometry ($i < 40^\circ$,
$-115^\circ < \xi < -75^\circ$).


\begin{figure}[h]
\begin{center}
\includegraphics{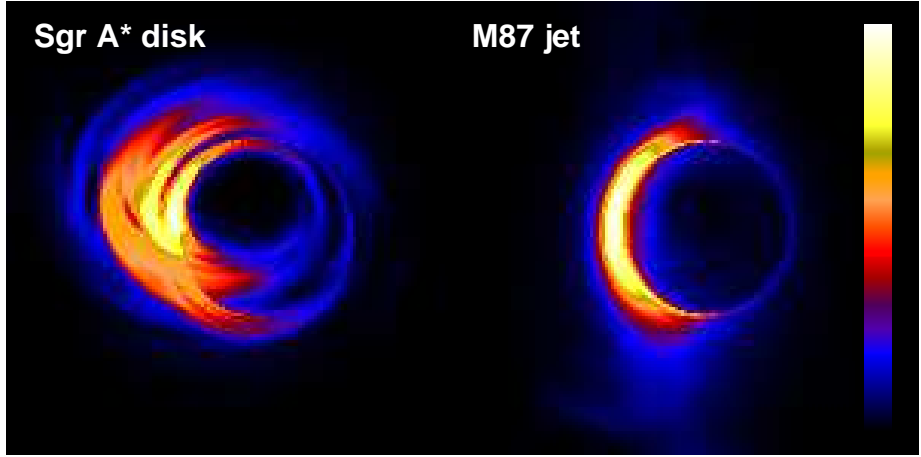}
\caption{\label{diskimg}False color fiducial images of an accretion disk model
  of Sgr A* (left) and a jet image of M87 (right). The left/right 
  asymmetries are caused by Doppler beaming from orbital motion, and the black hole shadow
  is clearly visible in the center of both images. The intensity scale
is linear with a dynamic range of 60, and the panels are each $12$ Schwarzschild radii ($\simeq100$ microarcseconds) across.}
\end{center}
\end{figure}

\section{Black Hole Images}

The millimeter emission from Sgr A* and M87 arises from the immediate
vicinity of the black hole ($r < 5-10$M for disk and $r <
2-4$M for jet emission). On these scales, relativistic effects tend to
dominate the appearance of the images, provided that the emission
region is not completely optically thick. The important effects are:
i) Doppler beaming from the component of the fluid velocity 
along the observer's line of sight, which is significant except for 
face-on viewing, and dominant for edge-on viewing; ii) light bending, which
causes the back of the accretion flow to appear above/below the black
hole (again except for face-on viewing), and iii) gravitational
lensing. Doppler beaming causes significant asymmetry between the
sides of the image where the gas is moving towards and away from the
observer. The combination of ii) and iii) importantly leads to the
black hole ``shadow'' \cite{bardeen1973,falcke,dexteretal2010}, the
transition between bound and unbound photon orbits. For lines of sight
inside the shadow, the emission comes from in front of the black hole,
where few photon emission angles escape to the observer at infinity. The gravitational lensing in the immediate vicinity often
leads to a bright ring at the circular photon orbit, surrounding the
shadow. Observing signatures of this shadow would constitute the first direct
evidence for a black hole event horizon. The combined effects of
asymmetry from Doppler beaming and light bending are typically both
important for favored models of M87 and Sgr A*, leading to characteristic
crescent image morphologies (an example is shown in the left panel of Figure
\ref{diskimg}). The shadow is also apparent in the image. 

\subsection{The Size of M87}

For the low inclination of M87, both disk and jet models also lead to
crescent images. In the jet case, the emission arises in the
\emph{counter-jet}, since the emission from the forward jet is along
lines of sight in the shadow and very little of it reaches the
observer (see the right panel of Figure \ref{diskimg} for an example). For this reason, we
can robustly predict the size of the jet 
launching region despite significant model uncertainties, as long as
the emission region is compact ($r \lesssim 10$M) and not completely
opaque ($\tau \lesssim 3$). We also assume the viewing geometry
estimated from observations of the large-scale jet.

We predict the size observed by mm-VLBI by interpolating fiducial
images to the locations sampled by the current mm-VLBI array
\cite{fishetal2011}. The results are $\simeq 33-44\mu$as, or $4-5$
Schwarzschild radii \cite{dexteretal2012}, very similar to existing
observations of Sgr A*. 

\begin{figure}[h]
\includegraphics[width=20pc]{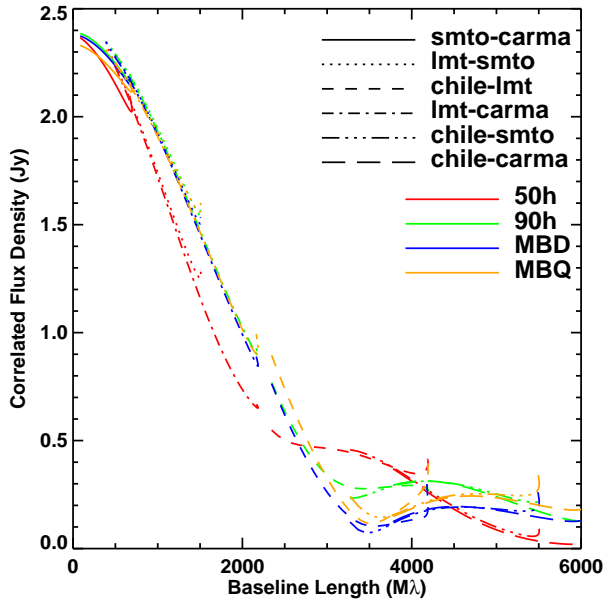}\hspace{2pc}
\begin{minipage}[b]{16pc}
\caption{\label{bhshadow}Visibility amplitude (Fourier transform of image intensity) as a function of baseline length for best fit models from \cite{dexteretal2010} at $1.3$mm. The telescopes considered are in Arizona (SMTO), California (CARMA), Mexico (LMT) and Chile (APEX/ASTE/ALMA). The BH shadow appears as the local minimum near 3000M$\lambda$ on the LMT-Chile and Chile-CARMA baselines.}
\end{minipage}
\end{figure}

\subsection{Black Hole Shadow}

Best fit models of Sgr A* and fiducial jet or disk/jet models of M87
all lead to crescent images, where Doppler beaming leads to
significant asymmetry but the front and back of the accretion flow
lead to significant contrast above and below the black hole. The black
hole shadow is then visible on orientations probing the front/back of
the accretion flow, while the image appears roughly Gaussian on
baselines 
probing orientations that are significantly affected by Doppler
beaming. For both sources, the models predict that the black hole
shadow is not visible with current telescopes (roughly East-West baselines), but may be observed
with North-South baselines including telescopes in Chile (e.g., ALMA) or Mexico
(LMT) (see Figure \ref{bhshadow}). These are also the telescopes with
the greatest ability to 
distinguish between and test the various models. Adding a nearly
orthogonal baseline orientation allows a probe of the 2D structure of the
image. There is also a
signature of the shadow in the closure phase \cite{dexteretal2010},
and combining amplitude and phase information probably offers the best 
chance for its detection.

\section{Variability}

These radiative models are the first to self-consistently include
variability from first principles. The variability in the accretion
disk emission is driven by fluctuations in the particle density and
magnetic field strength, which in turn are caused by magnetic
turbulence driven by the MRI \cite{dexter2009}. These fluctuations are
highly correlated with those in the accretion rate, although the power
law index of the power spectrum differs between the two quantities
\cite{noblekrolik2009,dexteretal2010}. 

\section{Discussion}

We have discussed the first models of the event horizon scale
millimeter-wavelength emission in Sgr A* and M87 from 3D GRMHD
simulations. The models are constructed by performing general
relativistic radiative transfer on the simulation data via ray
tracing, and can then be used to compare to or make predictions for
mm-VLBI observations. Even with limited data, the observations are
able to constrain the viewing geometry and physical conditions in the
Sgr A* accretion flow. Thus far, the uncertainties in the parameter estimates
decrease, while the values remain consistent, as more mm-VLBI data
become available. For M87, we predict a size of $33-44\mu$as,
where within that range the size depends on the total millimeter flux
of the source at the time of observation, whether the emission comes
from a disk or jet, and on the viewing geometry within the allowed
range. In both sources, the variability from the models agrees with
observations, especially in Sgr A* where even the morphology of the
millimeter flares can be reproduced with no additional free
parameters. In both sources, the black hole shadow, a signature of the
event horizon, may be accessible to future mm-VLBI observations using
ALMA and/or the LMT.

\subsection{Uncertainties}

The results presented here apply to existing simulations with modest
disk scale heights ($H/R \simeq 0.2$) at relatively high black hole
spins ($a \simeq 0.5-0.9375$) where the black hole spin and accretion
disk angular momentum axes are aligned. We expect the accretion flow
in low-luminosity sources should be somewhat thicker ($H/R \simeq 1$). The first
GRMHD simulations with $H/R \simeq 1$ have recently been carried out
(McKinney et al., in preparation), and may be used to study both Sgr
A* and M87. These simulations span a wide range of black hole spin and
initial magnetic field configuration, and will allow a fairly robust
parameter study of these radiative models including estimates of all
relevant parameters including the black hole spin.

For the accretion disk component, a constant ratio of $T_i/T_e$ is
assumed with no justification. This assumption may be okay, since the
millimeter emission region in these sources is so compact that their 
physical conditions do not vary greatly. However, this approximation
should be tested in the future. 

The primary uncertainty in the jet modeling is the dissipation physics
and jet mass loading, both of which are important for constructing
images. When the emission region is compact ($R 
\lesssim 5$M), relativistic effects tend to shape the images into
crescents for a large range of emission geometries. The jet emission
region may be extended, however, in which case most of the results for
jet images discussed here for $1.3$mm do not apply. This is true
observationally for M87 at 7mm. We have assumed the
non-thermal particle density is proportional to the magnetic energy
density, but other sub-grid prescriptions should be tried as well, if only to
gauge the range of viable jet image morphologies.

Finally, in geometrically thick sources the accretion flow and black
hole spin orientations are likely to remain misaligned. This disk
``tilt'' has drastic consequences for the structure and evolution of
the accretion flow \cite{fragile2007,fragile2008,fragiletilt2009} and
its observational properties \cite{dexterfragile2011}. The conclusions
reached here regarding crescent images, the visibility of black hole
shadows, and variability driven by MRI turbulence all apparently hold
for similar models from simulations with a $15^\circ$ tilt
\cite{dexter2011}. However, the parameter estimates for Sgr A*
discussed above only apply in the aligned case. 

\ack
This work was partially supported by NASA Earth \&
Space Science Fellowship NNX08AX59H (JD), STScI grant
HST-GO-11732.02-A, NSF grant AST 0807385, NASA grant 05-ATP05-96, and 
NASA Chandra Fellowship PF7-80048 (JCM). 

\section*{References}

\bibliographystyle{iopart-num}
\bibliography{master}

\end{document}